\documentclass[aps,prm,showpacs,floatfix,twocolumn,byrevtex,superscriptaddress]{revtex4-2}
\usepackage{amsmath}
\usepackage{amssymb}
\usepackage{amstext}
\usepackage{amsopn}
\usepackage{amsfonts}
\usepackage{amsxtra}
\usepackage[english]{babel}
\usepackage{graphicx}
\usepackage{float}
\usepackage{bm}
\usepackage{multirow}
\usepackage{dcolumn}
\usepackage{color}
\usepackage{hyperref}

\newcommand{\LPNO}{La$_{3-x}$Pr$_{x}$Ni$_{2}$O$_{7}$}
\newcommand{\LNO}{La$_{3}$Ni$_{2}$O$_{7-\delta}$}
\newcommand{\LSNO}{La$_{2.55}$Sr$_{0.45}$Ni$_{2}$O$_{7}$}
\newcommand{\LCNO}{La$_{2.55}$Ca$_{0.45}$Ni$_{2}$O$_{7}$}
\newcommand{\LANO}{La$_{3-x}$A$_{x}$Ni$_{2}$O$_{7}$}

\newcommand{\LaNiO}{La$_{4-x}$A$_{x}$Ni$_{3}$O$_{10}$}
\newcommand{\LAO}{LaAlO$_{3}$}
\newcommand{\targLNO}{La$_{1.7}$A$_{0.3}$NiO$_{4}$}

\begin{document}

\title{Decoupling between $d_{x^2-y^2}$ and $d_{z^2}$ orbitals in hole doped La$_3$Ni$_2$O$_7$}
\author{Yuecong Liu}
\author{Mengjun Ou}
\author{Yi Wang}
\author{Hai-Hu~Wen}
\email{hhwen@nju.edu.cn}
\affiliation{National Laboratory of Solid State Microstructures and Department of Physics, Collaborative Innovation Center of Advanced Microstructures, Nanjing University, Nanjing 210093, China}

\date{\today}
%
%

\begin{abstract}
Through Sr and Ca doping to the La sites, we successfully obtained the hole doped La$_{3-x}$A$_x$Ni$_2$O$_7$ (A = Sr and Ca) thin films by using the pulsed-laser deposition technique. Temperature dependent resistivity shows an upturn at low temperatures, but some clear instabilities, either due to structure or the releasing of strain between the film and substrate, occur at high temperatures. After annealing the films under high pressure of oxygen atmosphere, the upturn at low temperatures is strongly suppressed; the high temperature instability is completely removed. Hall effect measurements show a clear hole-charge carrier behavior with the carrier density of an order of magnitude higher compared with the undoped films. Surprisingly, it is found that the Hall coefficient is almost temperature independent in the whole temperature region, indicating the absence of multiband effect and suggesting the decoupling of the $d_{x^2-y^2}$ and $d_{z^2}$ orbitals. This is contradicting to the rigid band picture of a bonding $d_{z^2}$ band just below the Fermi energy in the pristine sample.
\end{abstract}

\maketitle

%
\section{INTRODUCTION}
%
Recently, superconductivity with a transition temperature around $T_c\approx$ 80~K in bilayer \LNO\ was reported under high pressure~\cite{sun2023signatures}. This is of great interest to the community, since high temperature superconductivity was long expected in nickelates~\cite{anisimov1999PRB, lee2004infinite, poltavets2010bulk, zhang2017large} and previously reported superconducting transition temperature in infinite-layers and quintuple-layer nickelates is only close to 20~K at ambient pressure and 30~K at high pressure~\cite{li2019superconductivity, osada2020superconducting, osada2021nickelate, zeng2022superconductivity, sun2023evidence, wang2022pressure, pan2022superconductivity}. Subsequently, zero resistance state in the \LNO\ was confirmed in both single crystals~\cite{zhang2024high, Hou2023CPL, zhou2024investigations, li2024pressuredriven} and polycrystalline samples~\cite{wang2023observation, wang2024pressure}. Up to now, superconductivity under high pressure was also found in Pr-doped \LPNO\ samples and trilayer Ruddlesden-Popper (RP) La$_{4}$Ni$_{3}$O$_{10}$ and Pr$_{4}$Ni$_{3}$O$_{10}$ ~\cite{zhu2024superconductivity, li2024signature, sakakibara2024theoretical, zhang2024superconductivity, li2024structural, huang2024signature, Wang2024CPL}. These results manifest that the nickelate superconductors are becoming the third unconventional superconducting family of high-$T_c$ superconductors after copper-oxide and iron-based superconductors. This is of great importance for the study of unconventional superconductivity mechanisms.

The bilayer \LNO\ has a triangle shape superconducting phase diagram under high pressure and its maximum superconducting volume fraction observed in experiment is about 48\%~\cite{li2024pressuredriven}. The pairing mechanism of \LNO\ is debatable~\cite{Zhang2023PRB, Luo2024npj, Yang2023PRB, Sakakibara2023PRL, Liu2023PRL, gu2023effective, Qin2023PRB, Lechermann2023PRB, Heier2024PRB, liu2023role, Jiang2024CPL, yang2023inter, Shen2023CPL, lu2024interlayer, qu2024PRL, Fan2024PRB, Ouyang2024hund}, but most model suggest that the pairing is induced by the magnetic superexchange effect. Theoretical calculations indicate that at ambient pressure, there are two bands crossing the Fermi level, and the low-energy states are mainly contributed by the Ni $d_{x^2-y^2}$ and in-plane O 2$p$ orbitals, as well as the Ni $d_{z^2}$. Since the strong inter-layer coupling through the inner apical oxygen, the energy bands formed by $d_{z^2}$ orbitals split into bonding and anti-bonding bands, locating at below and above the Fermi level, respectively. When pressure is applied, the electronic band structure changes and the $\sigma$-bonding band is brought to and even cross the Fermi level~\cite{sun2023signatures, luo2023bilayer}. Earlier studies have shown that the oxygen level strongly influences the transport properties of $\rm{La_{3}Ni_2O_7}$~\cite{taniguchi1995transport}. Energy-filtered multislice electron ptychography indicated that the oxygen vacancies are mainly located at the inner apical sites~\cite{Dong2024Nature} between the two neighboring NiO$_6$ octahedrons, which might be crucial for superconductivity~\cite{Liu2023PRL, Zhang2024PRB}. In addition, for all of the RP nickelates so far, superconductivity can only be driven under high pressure, and the high-pressure superconducting phases all have tetragonal structure, which might be necessary for the superconductivity. Thus, obtaining a tetragonal structure phase at lower or ambient pressure may be crucial for obtaining superconductivity. A feasible solution is to fabricate thin films and use the strain provided by the substrate and interface effect to change the rotations and orientations of the oxygen octahedral. However, it seems not easy to control this effect because the substrate can only apply strain to the film in the in-plane directions. And inappropriate biaxial strain may lead to the formation of impurity phases in the thin film~\cite{cui2023strain}. Another way is to use ions with a larger radius (like Sr$^{2+}$, Ba$^{2+}$, ect.) to replace some of the La$^{3+}$, thereby attenuating the rotations and orientations of the oxygen octahedral~\cite{zhang1994structure, xu2023pressure, Jiao20241354504}.

In this paper, by using pulsed-laser deposition (PLD) technique, we successfully obtained the hole doped \LSNO\ (LSNO) and \LCNO\ (LCNO) thin films with 15\% Sr and Ca doping to the La sites. Their temperature-dependent resistivity shows metallic behavior with an upturn at low temperatures and some instabilities at high temperatures in the pristine thin films. After annealing the films under high pressure of oxygen atmosphere, the upturn at low temperatures are strongly suppressed and the instabilities at high temperatures disappeared. The Hall effect measurements of the LSNO and LCNO thin films show a clear hole-charge carrier behavior and the carrier density is nearly an order of magnitude higher than that of the undoped La$_3$Ni$_2$O$_7$. Surprisingly, the Hall coefficient of the hole doped samples is almost temperature-independent in the whole temperature region, which indicates the absence of multiband effect and suggests the decoupling of the $d_{x^2-y^2}$ and $d_{z^2}$ orbitals. This may be related to the orbital selectivity of doped holes.

%
\section{EXPERIMENTAL DETAILS}

The La$_{2.55}$A$_{0.45}$Ni$_{2}$O$_{7}$ (A=Sr and Ca) thin films were grown on single-crystal (00$l$)-oriented \LAO\ substrates by pulsed-laser deposition (PLD) technique. The substrates were heated to $800\rm{^\circ C}$ at $10\rm{^\circ C/min}$ prior to growth. The oxygen partial pressure was kept at around 30 Pa during growth. The laser spot area is about 4.68~$\rm{mm^2}$, and the laser fluence and frequency are 2.07~$\rm{J/cm^2}$ and 2~Hz during the growth, respectively. After growth, the PLD chamber is filled with high purity oxygen gas until the partial pressure reaches 5$\times10^4$~Pa. Then the samples were cooled down to $450\rm{^\circ C}$ at $8\rm{^\circ C/min}$ held at $450\rm{^\circ C}$ for 1~h, which is followed by further cooling down to room temperature with the same cooling rate. The targets used for ablation was prepared by sol-gel method~\cite{poltavets2006oxygen} and they have NiO and \targLNO\ mixed phases. X-ray diffraction (XRD) measurements were performed by using a Bruker D8 Advanced diffractometer with Cu $\rm{K_{\alpha1}}$ radiation. Each sample was cut into several pieces for high pressure oxygen treatment. The samples were placed in a quartz tube and wrapped in aluminium foil and annealed for 6 hours at an oxygen pressure of 5~MPa, using a Kejing OTF-1200X high-temperature high-pressure tube furnace. Two annealing temperatures were selected ($450\rm{^\circ C}$ and $500\rm{^\circ C}$) for the annealing. Transport properties were measured by a physical property measurement system (PPMS, Quantum Design) with a magnetic field up to 9~T.

%

%
\section{RESULTS AND DISCUSSION}
%

\begin{figure}[tb]
\includegraphics[width=\columnwidth]{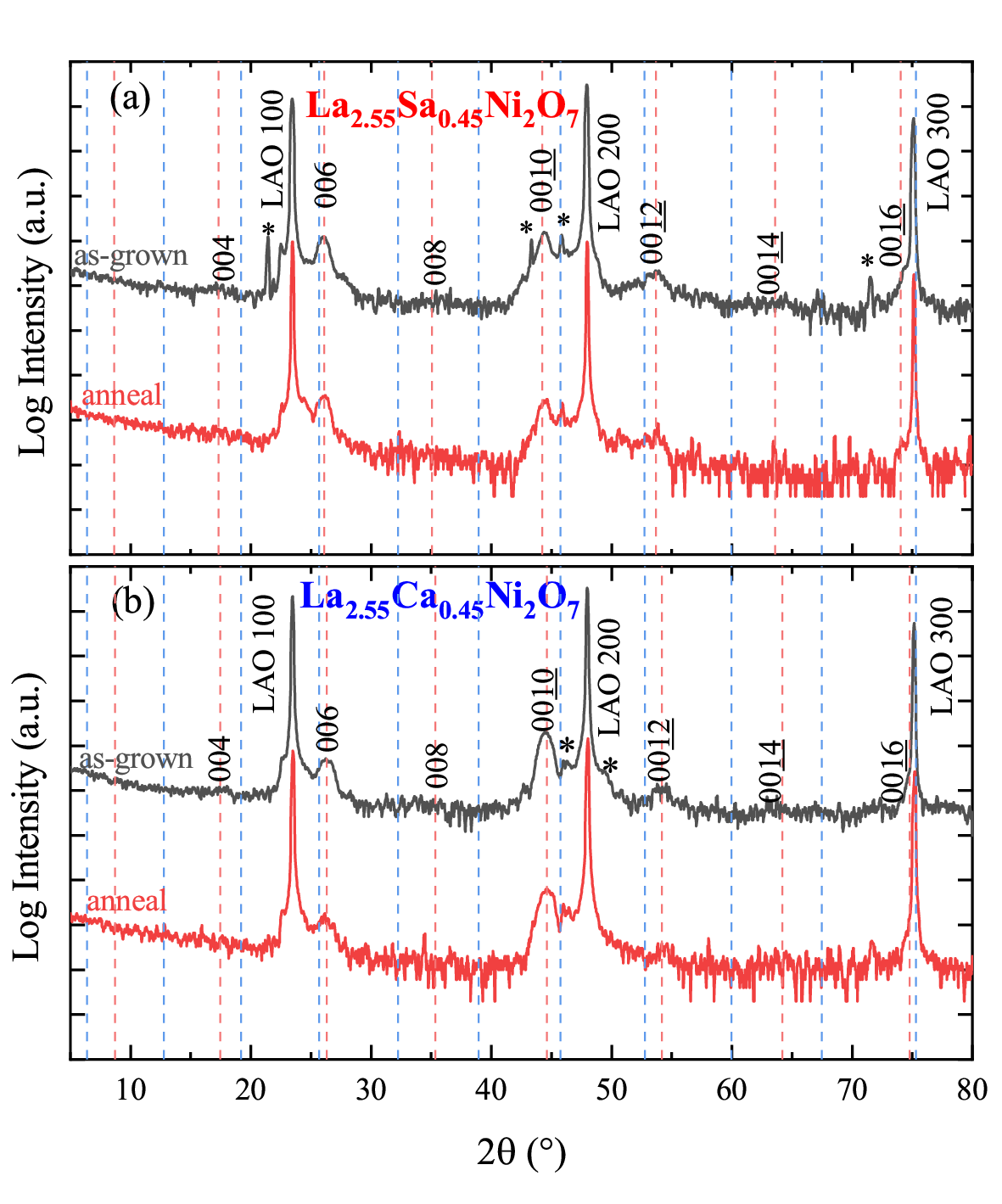}
\caption{Structural characterizations of the \LANO\ films. (a) XRD 2$\theta$-$\omega$ scans of the \LSNO\ thin film before and after annealing. (b) XRD 2$\theta$-$\omega$ scans of the \LCNO\ thin film before and after annealing. The annealing temperature for both samples is $500\rm{^\circ C}$. The dashed lines in red and blue represent the diffraction peak positions of \LANO\ and \LaNiO\ phases, respectively.}
\label{Figure1}
\end{figure}
Fig.~\ref{Figure1}(a) and (b) show the XRD 2$\theta$-$\omega$ scans of the \LANO\ thin films before and after annealing. It can be seen that the films are $c$-axis oriented. And the $c$-axis constants of the as-grown LSNO and LCNO thin films are 20.469$\pm$0.084 \r A and 20.299$\pm$0.036 \r A, respectively, which are slightly lower than the $c$-axis constants of the \LNO\ thin films on the same substrate~\cite{li2020epitaxial, liu2024growth}. The decrease in the $c$-axis constant of Ca-doped samples may be due to the fact that the ionic radius of Ca is smaller than that of La. However, the ion radius of strontium is larger than that of lanthanum but the $c$-axis lattice constant of the \LSNO\ film is also smaller than that of \LNO\ films. A similar phenomenon was also observed in polycrystalline samples~\cite{Jiao20241354504}, where the $a$ and $b$ axis constants increase with increasing Sr doping level, but simultaneously the $c$-axis lattice constant decreases. There is also a similar trend in La$_{3}$Ni$_{2-x}$Fe$_{x}$O$_{7}$~\cite{kiselev2019investigations}. But in an earlier study, the $c$-axis lattice constant increases with the increase of Sr doping level~\cite{zhang1994structure}.  We do not yet know what causes this difference, but we suspect it may be partially related to the oxygen content in the sample. After annealing, the $c$-axis lattice constants of \LSNO\ and \LCNO\ become 20.421$\pm$0.053 \r A and 20.312$\pm$0.084 \r A, respectively.
  
%
\begin{figure}[tb]
\includegraphics[width=\columnwidth]{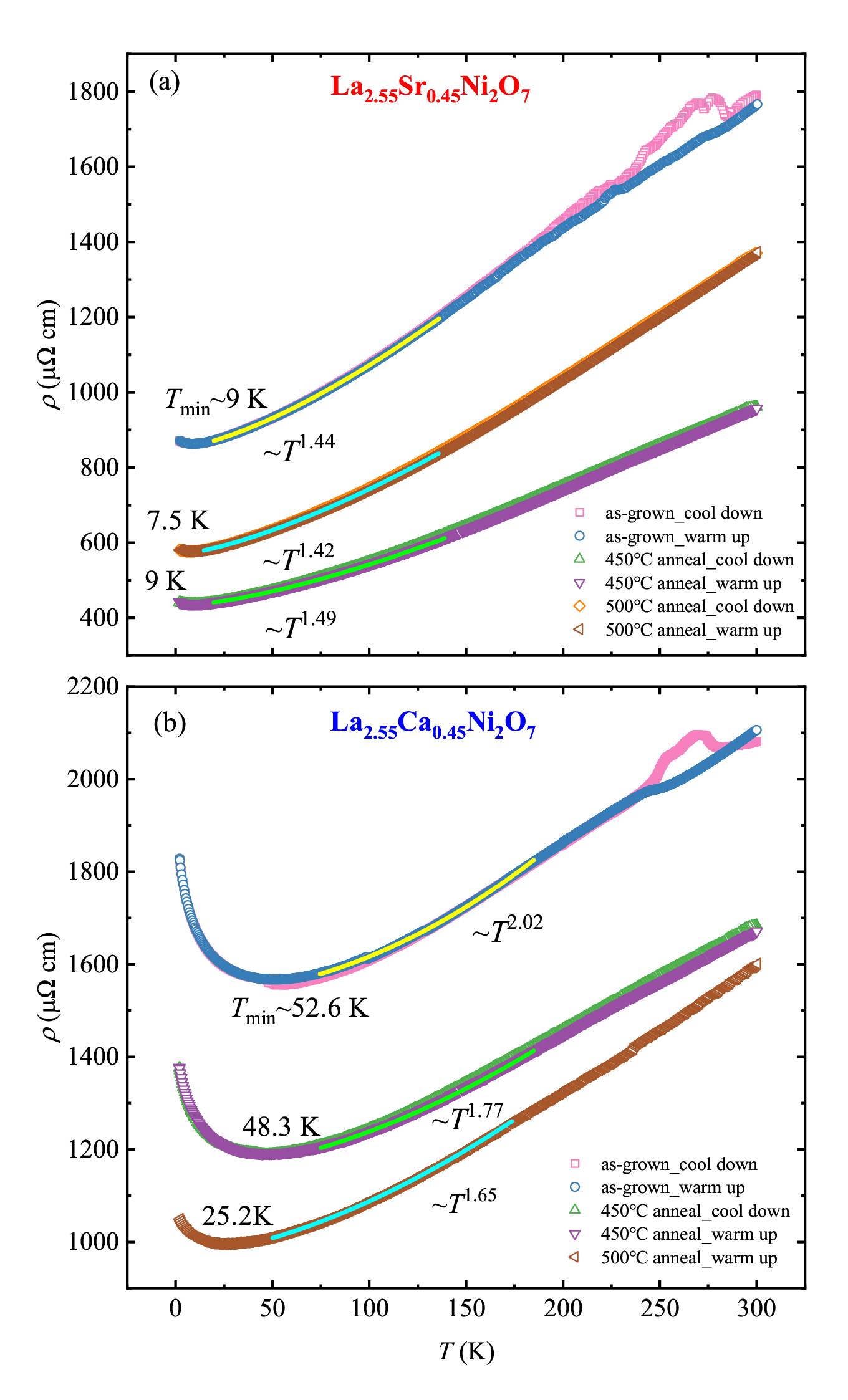}
\caption{$\rho$-T curves of the LSNO (a) and LCNO (b) thin films before and after anneal. The "warm up" and "cool down" in the legend represent the results from the measurement by rising or cooling temperature, respectively. The solid lines are the $\alpha$+$\beta$$T^{n}$ fits of the $\rho$-T at low temperatures.}
\label{Figure2}
\end{figure}
As shown in Fig.~\ref{Figure2}, the as-grown LSNO and LCNO thin films both show metallic behavior with an upturn at low temperatures. And the upturn of the LSNO thin film is much weaker than that of the LCNO and \LNO\ thin films, possibly because the larger Sr replaces part of the La and makes the nickel-oxygen surface flatter. After annealing the thin films at $500\rm{^\circ C}$ under high pressure of oxygen, the upturn is strongly suppressed and the temperature corresponding to the resistance minimum decreases. The solid lines in Fig.~\ref{Figure2} show the $\alpha$+$\beta$$T^{n}$ fits of the $\rho$-T at low temperatures, the fitted value of the exponential $n$ is labelled in the figure. It can be seen that the exponential $n$ of the LCNO thin films are all close to 2, indicating a Fermi liquid feature. But the $n$ value of the LSNO thin films is smaller, which may be an indication that the Sr doping enhances the correlation effect. The $\rho$-T curves during the cooling process of the as-grown LSNO and LCNO films both show some anomalies due to instabilities at high temperatures, the resistivity shows some hysteresis in the temperature ascending and descending processes; this hysteresis of resistivity disappears after annealing the films under high pressure of oxygen atmosphere (conducted later in a high pressure furnace with oxygen pressure of about 5 MPa). We think this instability of resistivity in the pristine films may be induced by a structural phase transition related to oxygen deficiency at high temperatures, or due to releasing the strain effect between the film and the substrate.

\begin{figure}[tb]
\includegraphics[width=\columnwidth]{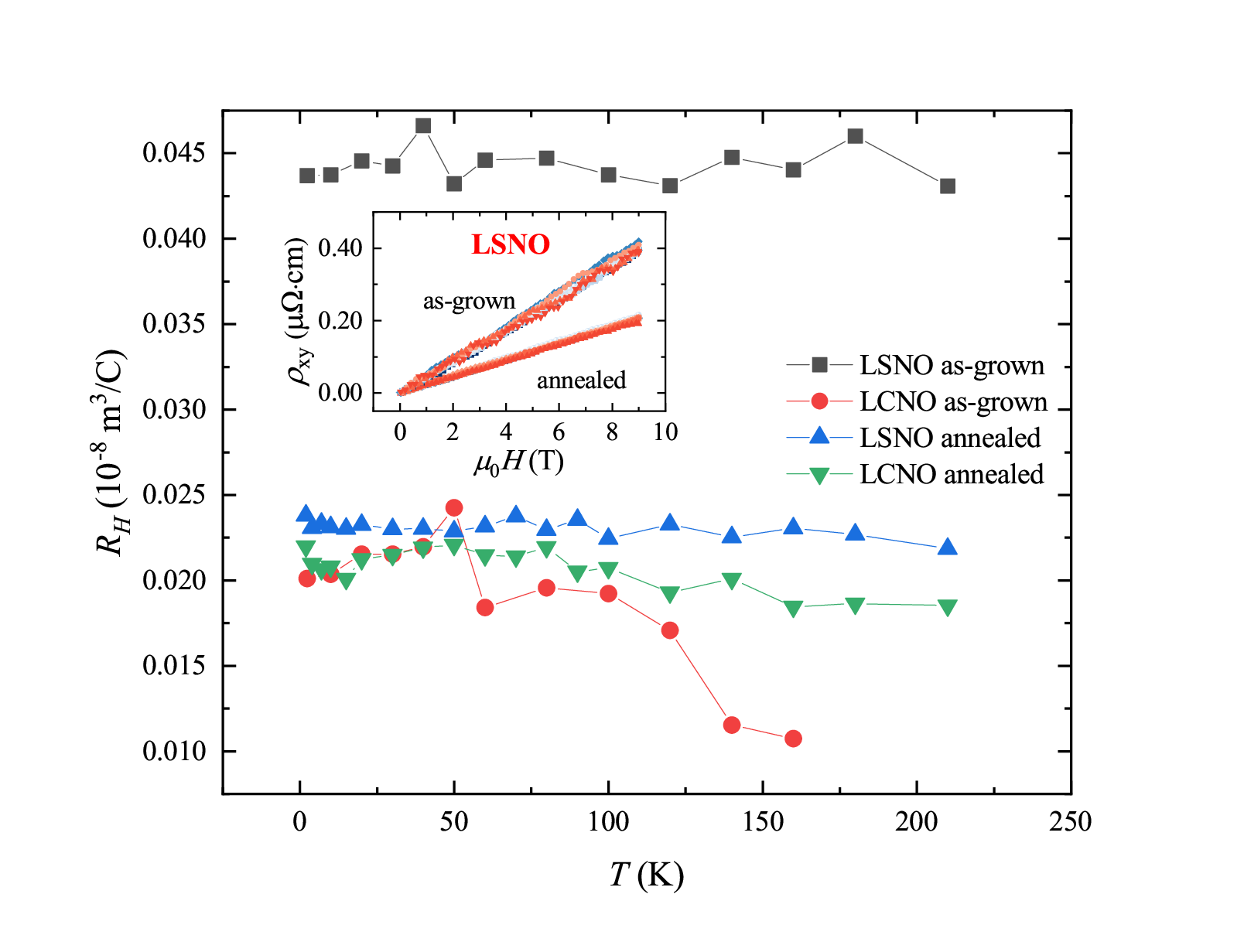}
\caption{Hall coefficients of the LSNO and LCNO thin films before and after annealing. The anneal temperature is $500\rm{^\circ C}$. The inset shows the magnetic field dependent $\rho_{xy}$ of the as-grown and annealed LSNO thin film at different temperatures.}
\label{Figure3}
\end{figure}
To understand the influence of hole doping and oxygen annealing on the electronic band structure, we measured the Hall effect before and after annealing of the samples. As shown in Fig.~\ref{Figure3}, all of the thin films show clear hole charge carrier behavior and the carrier density of the LSNO and LCNO thin films is nearly an order of magnitude higher than that of the undoped \LNO\ thin film~\cite{liu2024growth} and bulk samples~\cite{zhou2024evidence, taniguchi1995transport}. The Hall coefficient of the as-grown LSNO thin film is approximately twice that of the LCNO film and becomes similar to LCNO after oxygen annealing. Surprisingly, unlike the undoped \LNO\ in which the Hall coefficient has a strong temperature dependent, the Hall coefficient of the annealed \LANO\ and as-grown LSNO thin films is almost temperature-independent over the entire temperature range (up to 200 K). According to the rigid band model, the hole doping moves the Fermi surface down, then the $d_{z^2}$ bonding bands should be more closer or across the Fermi surface, the multiband effect should be more obvious, and even superconductivity may occur easily if the touching or crossing of the $d_{z^2}$ orbital with respect to the Fermi level takes place. However, the experiment results show the opposite. The temperature-independent Hall coefficient indicates the absence of multiband effect and suggests the decoupling of the $d_{x^2-y^2}$ and $d_{z^2}$ orbitals, which means that the doped holes may be mainly in the $d_{x^2-y^2}$ or the oxygen 2$p$ orbitals instead of the $d_{z^2}$ orbitals. This is probably due to the strong correlation effect, which makes it difficult for the holes to be doped into the $d_{z^2}$ orbital~\cite{Wu2024superexchange}. The inset of Fig.~\ref{Figure3} shows the magnetic field dependent transverse resistivity $\rho_{xy}$ of the as-grown and annealed LSNO thin film, which shows a good linear field dependence with positive slopes. And we can see that the $\rho_{xy}$ at different temperatures almost coincides with each other. For the LCNO thin film, the Hall coefficient of the as-grown sample shows some temperature dependence above 100 K, but becomes temperature independent over the whole temperature range after annealing. This may indicate that part of the doped holes occupy the oxygen 2$p$ orbitals.

\begin{figure}[htb]
	\includegraphics[width=\columnwidth]{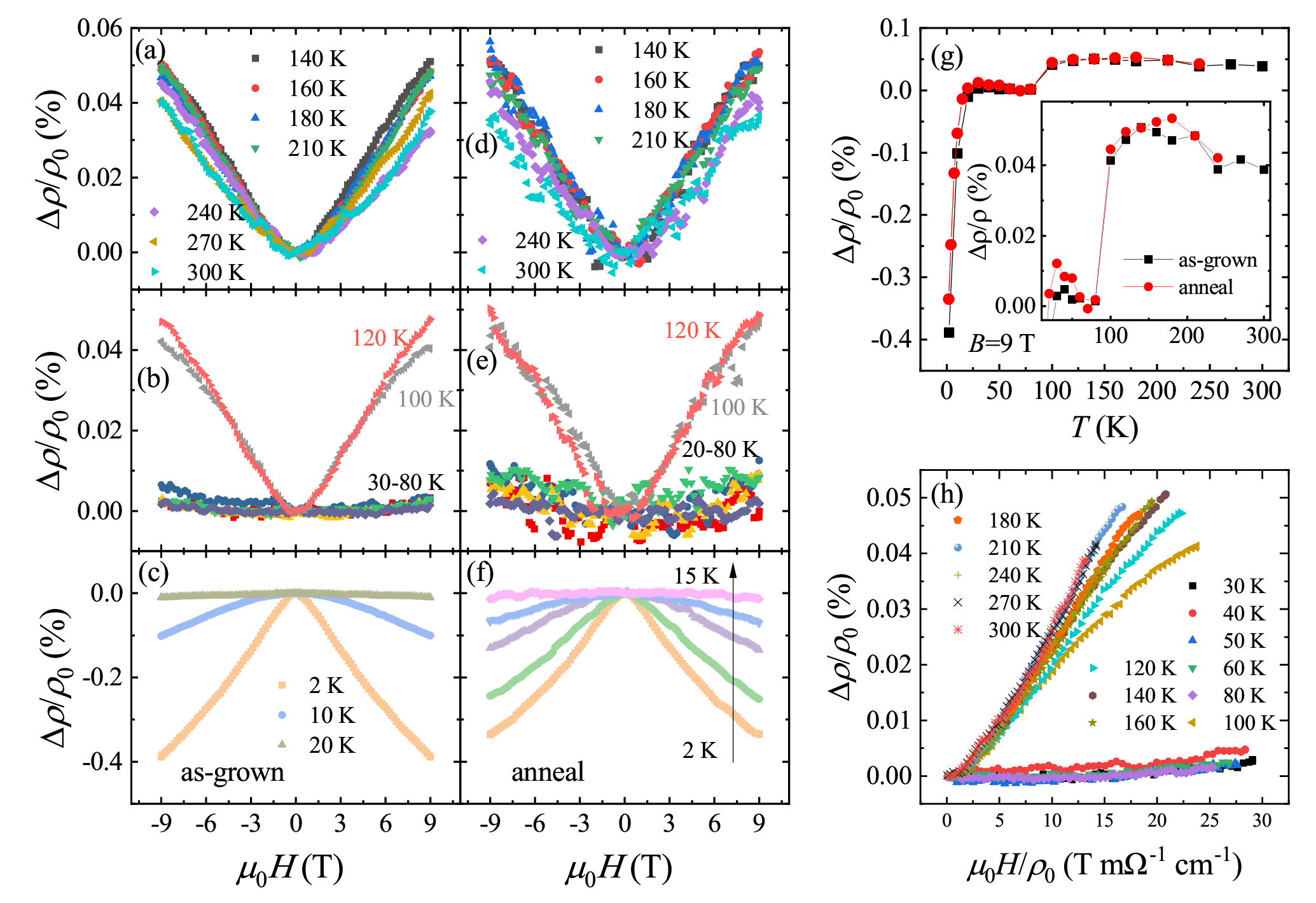}
	\caption{Magnetoresistance (MR) of the as-grown (a-c) and annealed (d-f) LSNO thin films. (g) Temperature dependent magnetoresistance of the as-grown LSNO thin film at 9 T. The inset shows a partial enlargement of (g). (h) The MR of as-grown LSNO thin film at temperatures above 30 K plotted by following Kohler's law.}
	\label{Figure4}
\end{figure}
The magnetoresistance (MR) defined as $\Delta\rho=\rho(H)-\rho_0$ (where $\rho(H)$ is the longitudinal resistivity in a magnetic field $H$ and $\rho_0$ is that at a zero-field) of the LSNO thin film has been systematically measured and the results are shown in Fig.~\ref{Figure4}(a)-(f), the magnetoresistance of the as-grown and annealed LSNO thin film is both negative at low temperatures and become positive at temperatures above about 30 K, which is similar to that of the undoped \LNO\ thin film. And after annealing under high pressure of oxygen atmosphere, the negative MR at low temperatures is slightly suppressed. The positive magnetoresistance of LSNO film is almost coincident at temperatures below 80 K, and there is a sharp increase at 80$\sim$100 K (as shown in Fig.~\ref{Figure4}(b) and (e)). We also plotted the temperature dependent MR of the as-grown LSNO thin film at 9 T in Fig.~\ref{Figure4}(g), the negative MR at low temperatures exhibits a rapid decrease with increasing temperature. The positive MR at around 30$\sim$80 K is very weak and the MR has a sharp increase at 80$\sim$100 K, rising from 0.005\% (average value of 30$\sim$80 K) to 0.045\%, changing by almost an order of magnitude. We have also observed similar phenomenon in metallic \LNO\ film~\cite{liu2024growth} as well as Pr doped \LPNO\ thin films (not shown here). It is unclear yet what causes this unusual increase in magnetoresistance. The MR of the as-grown LSNO thin film at temperatures above 30 K have been plotted by following the Kohler's law~\cite{ziman2001electrons}, namely $\Delta\rho/\rho_0=F(H\tau)=f(H/\rho_0)$ versus $H/\rho_0$ (where $F$ and $f$ represent some unknown functions) is shown in Fig.~\ref{Figure4}(h). It can be seen that Kohler's law is not obeyed in the LSNO thin films. This may be related to the abnormal increase of magnetoresistance at 80$\sim$100 K.

%
%

\section{SUMMARY}
By using the PLD technique, we successfully obtained the hole doped $\rm{La_{2.55}A_{0.45}Ni_2O_7}$ (A = Sr and Ca) thin films with 15\% Sr or Ca doping to the La sites. Both of the doped thin films have shorter $c$-axis lattice constants than that of the undoped $\rm{La_{3}Ni_2O_7}$ thin film. The temperature dependent resistivity shows metallic behavior with an upturn at low temperatures, and there are some clear instabilities at high temperatures. After annealing the films under high pressure of oxygen atmosphere, the instability disappears, the upturn at low temperatures is strongly suppressed. The Hall effect measurements show a clear hole-charge carrier behavior with the carrier density of nearly an order of magnitude higher than the undoped $\rm{La_{3}Ni_2O_7}$. And the Hall coefficient is almost temperature independent in the whole temperature region, indicating the absence of multiband contribution and suggesting the decoupling  of the $d_{x^2-y^2}$ and $d_{z^2}$ orbitals. This is contradicting to the rigid band picture of a bonding $d_{z^2}$ band just below the Fermi energy in the pristine sample. It is probably the strong correlation effect that makes it difficult for the holes to be doped into the Ni $d_{z^2}$ orbitals.
%
%

\begin{acknowledgments}
We acknowledge financial support from the National Key Research and Development Program of China  (No. 2022YFA1403201), National Natural Science Foundation of China (Nos. 11927809, 12061131001). 
\end{acknowledgments}



\begin{thebibliography}{57}
\expandafter\ifx\csname natexlab\endcsname\relax\def\natexlab#1{#1}\fi
\expandafter\ifx\csname bibnamefont\endcsname\relax
  \def\bibnamefont#1{#1}\fi
\expandafter\ifx\csname bibfnamefont\endcsname\relax
  \def\bibfnamefont#1{#1}\fi
\expandafter\ifx\csname citenamefont\endcsname\relax
  \def\citenamefont#1{#1}\fi
\expandafter\ifx\csname url\endcsname\relax
  \def\url#1{\texttt{#1}}\fi
\expandafter\ifx\csname urlprefix\endcsname\relax\def\urlprefix{URL }\fi
\providecommand{\bibinfo}[2]{#2}
\providecommand{\eprint}[2][]{\url{#2}}

\bibitem[{\citenamefont{Sun et~al.}(2023{\natexlab{a}})\citenamefont{Sun, Huo,
  Hu, Li, Liu, Han, Tang, Mao, Yang, Wang et~al.}}]{sun2023signatures}
\bibinfo{author}{\bibfnamefont{H.}~\bibnamefont{Sun}},
  \bibinfo{author}{\bibfnamefont{M.}~\bibnamefont{Huo}},
  \bibinfo{author}{\bibfnamefont{X.}~\bibnamefont{Hu}},
  \bibinfo{author}{\bibfnamefont{J.}~\bibnamefont{Li}},
  \bibinfo{author}{\bibfnamefont{Z.}~\bibnamefont{Liu}},
  \bibinfo{author}{\bibfnamefont{Y.}~\bibnamefont{Han}},
  \bibinfo{author}{\bibfnamefont{L.}~\bibnamefont{Tang}},
  \bibinfo{author}{\bibfnamefont{Z.}~\bibnamefont{Mao}},
  \bibinfo{author}{\bibfnamefont{P.}~\bibnamefont{Yang}},
  \bibinfo{author}{\bibfnamefont{B.}~\bibnamefont{Wang}}, \bibnamefont{et~al.},
  \bibinfo{journal}{Nature} \textbf{\bibinfo{volume}{621}},
  \bibinfo{pages}{493} (\bibinfo{year}{2023}{\natexlab{a}}).

\bibitem[{\citenamefont{Anisimov et~al.}(1999)\citenamefont{Anisimov,
  Bukhvalov, and Rice}}]{anisimov1999PRB}
\bibinfo{author}{\bibfnamefont{V.}~\bibnamefont{Anisimov}},
  \bibinfo{author}{\bibfnamefont{D.}~\bibnamefont{Bukhvalov}},
  \bibnamefont{and} \bibinfo{author}{\bibfnamefont{T.}~\bibnamefont{Rice}},
  \bibinfo{journal}{Phys. Rev. B} \textbf{\bibinfo{volume}{59}},
  \bibinfo{pages}{7901} (\bibinfo{year}{1999}).

\bibitem[{\citenamefont{Lee and Pickett}(2004)}]{lee2004infinite}
\bibinfo{author}{\bibfnamefont{K.-W.} \bibnamefont{Lee}} \bibnamefont{and}
  \bibinfo{author}{\bibfnamefont{W.}~\bibnamefont{Pickett}},
  \bibinfo{journal}{Phys. Rev. B} \textbf{\bibinfo{volume}{70}},
  \bibinfo{pages}{165109} (\bibinfo{year}{2004}).

\bibitem[{\citenamefont{Poltavets et~al.}(2010)\citenamefont{Poltavets,
  Lokshin, Nevidomskyy, Croft, Tyson, Hadermann, Van~Tendeloo, Egami, Kotliar,
  ApRoberts-Warren et~al.}}]{poltavets2010bulk}
\bibinfo{author}{\bibfnamefont{V.~V.} \bibnamefont{Poltavets}},
  \bibinfo{author}{\bibfnamefont{K.~A.} \bibnamefont{Lokshin}},
  \bibinfo{author}{\bibfnamefont{A.~H.} \bibnamefont{Nevidomskyy}},
  \bibinfo{author}{\bibfnamefont{M.}~\bibnamefont{Croft}},
  \bibinfo{author}{\bibfnamefont{T.~A.} \bibnamefont{Tyson}},
  \bibinfo{author}{\bibfnamefont{J.}~\bibnamefont{Hadermann}},
  \bibinfo{author}{\bibfnamefont{G.}~\bibnamefont{Van~Tendeloo}},
  \bibinfo{author}{\bibfnamefont{T.}~\bibnamefont{Egami}},
  \bibinfo{author}{\bibfnamefont{G.}~\bibnamefont{Kotliar}},
  \bibinfo{author}{\bibfnamefont{N.}~\bibnamefont{ApRoberts-Warren}},
  \bibnamefont{et~al.}, \bibinfo{journal}{Phys. Rev. Lett}
  \textbf{\bibinfo{volume}{104}}, \bibinfo{pages}{206403}
  (\bibinfo{year}{2010}).

\bibitem[{\citenamefont{Zhang et~al.}(2017)\citenamefont{Zhang, Botana,
  Freeland, Phelan, Zheng, Pardo, Norman, and Mitchell}}]{zhang2017large}
\bibinfo{author}{\bibfnamefont{J.}~\bibnamefont{Zhang}},
  \bibinfo{author}{\bibfnamefont{A.}~\bibnamefont{Botana}},
  \bibinfo{author}{\bibfnamefont{J.}~\bibnamefont{Freeland}},
  \bibinfo{author}{\bibfnamefont{D.}~\bibnamefont{Phelan}},
  \bibinfo{author}{\bibfnamefont{H.}~\bibnamefont{Zheng}},
  \bibinfo{author}{\bibfnamefont{V.}~\bibnamefont{Pardo}},
  \bibinfo{author}{\bibfnamefont{M.}~\bibnamefont{Norman}}, \bibnamefont{and}
  \bibinfo{author}{\bibfnamefont{J.}~\bibnamefont{Mitchell}},
  \bibinfo{journal}{Nat. Phys.} \textbf{\bibinfo{volume}{13}},
  \bibinfo{pages}{864} (\bibinfo{year}{2017}).

\bibitem[{\citenamefont{Li et~al.}(2019)\citenamefont{Li, Lee, Wang, Osada,
  Crossley, Lee, Cui, Hikita, and Hwang}}]{li2019superconductivity}
\bibinfo{author}{\bibfnamefont{D.}~\bibnamefont{Li}},
  \bibinfo{author}{\bibfnamefont{K.}~\bibnamefont{Lee}},
  \bibinfo{author}{\bibfnamefont{B.~Y.} \bibnamefont{Wang}},
  \bibinfo{author}{\bibfnamefont{M.}~\bibnamefont{Osada}},
  \bibinfo{author}{\bibfnamefont{S.}~\bibnamefont{Crossley}},
  \bibinfo{author}{\bibfnamefont{H.~R.} \bibnamefont{Lee}},
  \bibinfo{author}{\bibfnamefont{Y.}~\bibnamefont{Cui}},
  \bibinfo{author}{\bibfnamefont{Y.}~\bibnamefont{Hikita}}, \bibnamefont{and}
  \bibinfo{author}{\bibfnamefont{H.~Y.} \bibnamefont{Hwang}},
  \bibinfo{journal}{Nature} \textbf{\bibinfo{volume}{572}},
  \bibinfo{pages}{624} (\bibinfo{year}{2019}).

\bibitem[{\citenamefont{Osada et~al.}(2020)\citenamefont{Osada, Wang, Goodge,
  Lee, Yoon, Sakuma, Li, Miura, Kourkoutis, and
  Hwang}}]{osada2020superconducting}
\bibinfo{author}{\bibfnamefont{M.}~\bibnamefont{Osada}},
  \bibinfo{author}{\bibfnamefont{B.~Y.} \bibnamefont{Wang}},
  \bibinfo{author}{\bibfnamefont{B.~H.} \bibnamefont{Goodge}},
  \bibinfo{author}{\bibfnamefont{K.}~\bibnamefont{Lee}},
  \bibinfo{author}{\bibfnamefont{H.}~\bibnamefont{Yoon}},
  \bibinfo{author}{\bibfnamefont{K.}~\bibnamefont{Sakuma}},
  \bibinfo{author}{\bibfnamefont{D.}~\bibnamefont{Li}},
  \bibinfo{author}{\bibfnamefont{M.}~\bibnamefont{Miura}},
  \bibinfo{author}{\bibfnamefont{L.~F.} \bibnamefont{Kourkoutis}},
  \bibnamefont{and} \bibinfo{author}{\bibfnamefont{H.~Y.} \bibnamefont{Hwang}},
  \bibinfo{journal}{Nano Lett} \textbf{\bibinfo{volume}{20}},
  \bibinfo{pages}{5735} (\bibinfo{year}{2020}).

\bibitem[{\citenamefont{Osada et~al.}(2021)\citenamefont{Osada, Wang, Goodge,
  Harvey, Lee, Li, Kourkoutis, and Hwang}}]{osada2021nickelate}
\bibinfo{author}{\bibfnamefont{M.}~\bibnamefont{Osada}},
  \bibinfo{author}{\bibfnamefont{B.~Y.} \bibnamefont{Wang}},
  \bibinfo{author}{\bibfnamefont{B.~H.} \bibnamefont{Goodge}},
  \bibinfo{author}{\bibfnamefont{S.~P.} \bibnamefont{Harvey}},
  \bibinfo{author}{\bibfnamefont{K.}~\bibnamefont{Lee}},
  \bibinfo{author}{\bibfnamefont{D.}~\bibnamefont{Li}},
  \bibinfo{author}{\bibfnamefont{L.~F.} \bibnamefont{Kourkoutis}},
  \bibnamefont{and} \bibinfo{author}{\bibfnamefont{H.~Y.} \bibnamefont{Hwang}},
  \bibinfo{journal}{Adv. Mater} \textbf{\bibinfo{volume}{33}},
  \bibinfo{pages}{2104083} (\bibinfo{year}{2021}).

\bibitem[{\citenamefont{Zeng et~al.}(2022)\citenamefont{Zeng, Li, Chow, Cao,
  Zhang, Tang, Yin, Lim, Hu, Yang et~al.}}]{zeng2022superconductivity}
\bibinfo{author}{\bibfnamefont{S.}~\bibnamefont{Zeng}},
  \bibinfo{author}{\bibfnamefont{C.}~\bibnamefont{Li}},
  \bibinfo{author}{\bibfnamefont{L.~E.} \bibnamefont{Chow}},
  \bibinfo{author}{\bibfnamefont{Y.}~\bibnamefont{Cao}},
  \bibinfo{author}{\bibfnamefont{Z.}~\bibnamefont{Zhang}},
  \bibinfo{author}{\bibfnamefont{C.~S.} \bibnamefont{Tang}},
  \bibinfo{author}{\bibfnamefont{X.}~\bibnamefont{Yin}},
  \bibinfo{author}{\bibfnamefont{Z.~S.} \bibnamefont{Lim}},
  \bibinfo{author}{\bibfnamefont{J.}~\bibnamefont{Hu}},
  \bibinfo{author}{\bibfnamefont{P.}~\bibnamefont{Yang}}, \bibnamefont{et~al.},
  \bibinfo{journal}{Sci. Adv} \textbf{\bibinfo{volume}{8}},
  \bibinfo{pages}{eabl9927} (\bibinfo{year}{2022}).

\bibitem[{\citenamefont{Sun et~al.}(2023{\natexlab{b}})\citenamefont{Sun, Li,
  Liu, Yang, Li, Wei, Jin, Yan, Sun, Guo et~al.}}]{sun2023evidence}
\bibinfo{author}{\bibfnamefont{W.}~\bibnamefont{Sun}},
  \bibinfo{author}{\bibfnamefont{Y.}~\bibnamefont{Li}},
  \bibinfo{author}{\bibfnamefont{R.}~\bibnamefont{Liu}},
  \bibinfo{author}{\bibfnamefont{J.}~\bibnamefont{Yang}},
  \bibinfo{author}{\bibfnamefont{J.}~\bibnamefont{Li}},
  \bibinfo{author}{\bibfnamefont{W.}~\bibnamefont{Wei}},
  \bibinfo{author}{\bibfnamefont{G.}~\bibnamefont{Jin}},
  \bibinfo{author}{\bibfnamefont{S.}~\bibnamefont{Yan}},
  \bibinfo{author}{\bibfnamefont{H.}~\bibnamefont{Sun}},
  \bibinfo{author}{\bibfnamefont{W.}~\bibnamefont{Guo}}, \bibnamefont{et~al.},
  \bibinfo{journal}{Adv. Mater} \textbf{\bibinfo{volume}{35}},
  \bibinfo{pages}{2303400} (\bibinfo{year}{2023}{\natexlab{b}}).

\bibitem[{\citenamefont{Wang et~al.}(2022)\citenamefont{Wang, Yang, Yang, Chen,
  Zhang, Zhang, Zhu, Uwatoko, Gu, Dong et~al.}}]{wang2022pressure}
\bibinfo{author}{\bibfnamefont{N.}~\bibnamefont{Wang}},
  \bibinfo{author}{\bibfnamefont{M.}~\bibnamefont{Yang}},
  \bibinfo{author}{\bibfnamefont{Z.}~\bibnamefont{Yang}},
  \bibinfo{author}{\bibfnamefont{K.}~\bibnamefont{Chen}},
  \bibinfo{author}{\bibfnamefont{H.}~\bibnamefont{Zhang}},
  \bibinfo{author}{\bibfnamefont{Q.}~\bibnamefont{Zhang}},
  \bibinfo{author}{\bibfnamefont{Z.}~\bibnamefont{Zhu}},
  \bibinfo{author}{\bibfnamefont{Y.}~\bibnamefont{Uwatoko}},
  \bibinfo{author}{\bibfnamefont{L.}~\bibnamefont{Gu}},
  \bibinfo{author}{\bibfnamefont{X.}~\bibnamefont{Dong}}, \bibnamefont{et~al.},
  \bibinfo{journal}{Nat. Commun} \textbf{\bibinfo{volume}{13}},
  \bibinfo{pages}{4367} (\bibinfo{year}{2022}).

\bibitem[{\citenamefont{Pan et~al.}(2022)\citenamefont{Pan, Ferenc~Segedin,
  LaBollita, Song, Nica, Goodge, Pierce, Doyle, Novakov, C{\'o}rdova~Carrizales
  et~al.}}]{pan2022superconductivity}
\bibinfo{author}{\bibfnamefont{G.~A.} \bibnamefont{Pan}},
  \bibinfo{author}{\bibfnamefont{D.}~\bibnamefont{Ferenc~Segedin}},
  \bibinfo{author}{\bibfnamefont{H.}~\bibnamefont{LaBollita}},
  \bibinfo{author}{\bibfnamefont{Q.}~\bibnamefont{Song}},
  \bibinfo{author}{\bibfnamefont{E.~M.} \bibnamefont{Nica}},
  \bibinfo{author}{\bibfnamefont{B.~H.} \bibnamefont{Goodge}},
  \bibinfo{author}{\bibfnamefont{A.~T.} \bibnamefont{Pierce}},
  \bibinfo{author}{\bibfnamefont{S.}~\bibnamefont{Doyle}},
  \bibinfo{author}{\bibfnamefont{S.}~\bibnamefont{Novakov}},
  \bibinfo{author}{\bibfnamefont{D.}~\bibnamefont{C{\'o}rdova~Carrizales}},
  \bibnamefont{et~al.}, \bibinfo{journal}{Nat. Mater}
  \textbf{\bibinfo{volume}{21}}, \bibinfo{pages}{160} (\bibinfo{year}{2022}).

\bibitem[{\citenamefont{Zhang et~al.}(2024{\natexlab{a}})\citenamefont{Zhang,
  Su, Huang, Shan, Sun, Huo, Ye, Zhang, Yang, Xu et~al.}}]{zhang2024high}
\bibinfo{author}{\bibfnamefont{Y.}~\bibnamefont{Zhang}},
  \bibinfo{author}{\bibfnamefont{D.}~\bibnamefont{Su}},
  \bibinfo{author}{\bibfnamefont{Y.}~\bibnamefont{Huang}},
  \bibinfo{author}{\bibfnamefont{Z.}~\bibnamefont{Shan}},
  \bibinfo{author}{\bibfnamefont{H.}~\bibnamefont{Sun}},
  \bibinfo{author}{\bibfnamefont{M.}~\bibnamefont{Huo}},
  \bibinfo{author}{\bibfnamefont{K.}~\bibnamefont{Ye}},
  \bibinfo{author}{\bibfnamefont{J.}~\bibnamefont{Zhang}},
  \bibinfo{author}{\bibfnamefont{Z.}~\bibnamefont{Yang}},
  \bibinfo{author}{\bibfnamefont{Y.}~\bibnamefont{Xu}}, \bibnamefont{et~al.},
  \bibinfo{journal}{Nat. Phys} pp. \bibinfo{pages}{1--5}
  (\bibinfo{year}{2024}{\natexlab{a}}).

\bibitem[{\citenamefont{Hou et~al.}(2023)\citenamefont{Hou, Yang, Liu, Li,
  Shan, Ma, Wang, Wang, Guo, Sun et~al.}}]{Hou2023CPL}
\bibinfo{author}{\bibfnamefont{J.}~\bibnamefont{Hou}},
  \bibinfo{author}{\bibfnamefont{P.-T.} \bibnamefont{Yang}},
  \bibinfo{author}{\bibfnamefont{Z.-Y.} \bibnamefont{Liu}},
  \bibinfo{author}{\bibfnamefont{J.-Y.} \bibnamefont{Li}},
  \bibinfo{author}{\bibfnamefont{P.-F.} \bibnamefont{Shan}},
  \bibinfo{author}{\bibfnamefont{L.}~\bibnamefont{Ma}},
  \bibinfo{author}{\bibfnamefont{G.}~\bibnamefont{Wang}},
  \bibinfo{author}{\bibfnamefont{N.-N.} \bibnamefont{Wang}},
  \bibinfo{author}{\bibfnamefont{H.-Z.} \bibnamefont{Guo}},
  \bibinfo{author}{\bibfnamefont{J.-P.} \bibnamefont{Sun}},
  \bibnamefont{et~al.}, \bibinfo{journal}{Chinese Physics Letters}
  \textbf{\bibinfo{volume}{40}}, \bibinfo{pages}{117302}
  (\bibinfo{year}{2023}).

\bibitem[{\citenamefont{Zhou et~al.}(2024{\natexlab{a}})\citenamefont{Zhou,
  Guo, Cai, Sun, Wang, Zhao, Han, Chen, Chen, Wu
  et~al.}}]{zhou2024investigations}
\bibinfo{author}{\bibfnamefont{Y.}~\bibnamefont{Zhou}},
  \bibinfo{author}{\bibfnamefont{J.}~\bibnamefont{Guo}},
  \bibinfo{author}{\bibfnamefont{S.}~\bibnamefont{Cai}},
  \bibinfo{author}{\bibfnamefont{H.}~\bibnamefont{Sun}},
  \bibinfo{author}{\bibfnamefont{P.}~\bibnamefont{Wang}},
  \bibinfo{author}{\bibfnamefont{J.}~\bibnamefont{Zhao}},
  \bibinfo{author}{\bibfnamefont{J.}~\bibnamefont{Han}},
  \bibinfo{author}{\bibfnamefont{X.}~\bibnamefont{Chen}},
  \bibinfo{author}{\bibfnamefont{Y.}~\bibnamefont{Chen}},
  \bibinfo{author}{\bibfnamefont{Q.}~\bibnamefont{Wu}}, \bibnamefont{et~al.}
  (\bibinfo{year}{2024}{\natexlab{a}}), \eprint{arXiv:2311.12361}.

\bibitem[{\citenamefont{Li et~al.}(2024{\natexlab{a}})\citenamefont{Li, Ma,
  Zhang, Huang, Huang, Huo, Hu, Dong, He, Liao et~al.}}]{li2024pressuredriven}
\bibinfo{author}{\bibfnamefont{J.}~\bibnamefont{Li}},
  \bibinfo{author}{\bibfnamefont{P.}~\bibnamefont{Ma}},
  \bibinfo{author}{\bibfnamefont{H.}~\bibnamefont{Zhang}},
  \bibinfo{author}{\bibfnamefont{X.}~\bibnamefont{Huang}},
  \bibinfo{author}{\bibfnamefont{C.}~\bibnamefont{Huang}},
  \bibinfo{author}{\bibfnamefont{M.}~\bibnamefont{Huo}},
  \bibinfo{author}{\bibfnamefont{D.}~\bibnamefont{Hu}},
  \bibinfo{author}{\bibfnamefont{Z.}~\bibnamefont{Dong}},
  \bibinfo{author}{\bibfnamefont{C.}~\bibnamefont{He}},
  \bibinfo{author}{\bibfnamefont{J.}~\bibnamefont{Liao}}, \bibnamefont{et~al.}
  (\bibinfo{year}{2024}{\natexlab{a}}), \eprint{arXiv:2404.11369}.

\bibitem[{\citenamefont{Wang et~al.}(2023)\citenamefont{Wang, Wang, Wang, Shi,
  Shen, Hou, Ma, Yang, Liu, Zhang et~al.}}]{wang2023observation}
\bibinfo{author}{\bibfnamefont{G.}~\bibnamefont{Wang}},
  \bibinfo{author}{\bibfnamefont{N.}~\bibnamefont{Wang}},
  \bibinfo{author}{\bibfnamefont{Y.}~\bibnamefont{Wang}},
  \bibinfo{author}{\bibfnamefont{L.}~\bibnamefont{Shi}},
  \bibinfo{author}{\bibfnamefont{X.}~\bibnamefont{Shen}},
  \bibinfo{author}{\bibfnamefont{J.}~\bibnamefont{Hou}},
  \bibinfo{author}{\bibfnamefont{H.}~\bibnamefont{Ma}},
  \bibinfo{author}{\bibfnamefont{P.}~\bibnamefont{Yang}},
  \bibinfo{author}{\bibfnamefont{Z.}~\bibnamefont{Liu}},
  \bibinfo{author}{\bibfnamefont{H.}~\bibnamefont{Zhang}}, \bibnamefont{et~al.}
  (\bibinfo{year}{2023}), \eprint{arXiv:2311.08212}.

\bibitem[{\citenamefont{Wang et~al.}(2024{\natexlab{a}})\citenamefont{Wang,
  Wang, Shen, Hou, Ma, Shi, Ren, Gu, Ma, Yang et~al.}}]{wang2024pressure}
\bibinfo{author}{\bibfnamefont{G.}~\bibnamefont{Wang}},
  \bibinfo{author}{\bibfnamefont{N.}~\bibnamefont{Wang}},
  \bibinfo{author}{\bibfnamefont{X.}~\bibnamefont{Shen}},
  \bibinfo{author}{\bibfnamefont{J.}~\bibnamefont{Hou}},
  \bibinfo{author}{\bibfnamefont{L.}~\bibnamefont{Ma}},
  \bibinfo{author}{\bibfnamefont{L.}~\bibnamefont{Shi}},
  \bibinfo{author}{\bibfnamefont{Z.}~\bibnamefont{Ren}},
  \bibinfo{author}{\bibfnamefont{Y.}~\bibnamefont{Gu}},
  \bibinfo{author}{\bibfnamefont{H.}~\bibnamefont{Ma}},
  \bibinfo{author}{\bibfnamefont{P.}~\bibnamefont{Yang}}, \bibnamefont{et~al.},
  \bibinfo{journal}{Phys. Rev. X} \textbf{\bibinfo{volume}{14}},
  \bibinfo{pages}{011040} (\bibinfo{year}{2024}{\natexlab{a}}).

\bibitem[{\citenamefont{Zhu et~al.}(2024)\citenamefont{Zhu, Zhang, Pan, Chen,
  Peng, Chen, Ren, Liu, Li, Xing et~al.}}]{zhu2024superconductivity}
\bibinfo{author}{\bibfnamefont{Y.}~\bibnamefont{Zhu}},
  \bibinfo{author}{\bibfnamefont{E.}~\bibnamefont{Zhang}},
  \bibinfo{author}{\bibfnamefont{B.}~\bibnamefont{Pan}},
  \bibinfo{author}{\bibfnamefont{X.}~\bibnamefont{Chen}},
  \bibinfo{author}{\bibfnamefont{D.}~\bibnamefont{Peng}},
  \bibinfo{author}{\bibfnamefont{L.}~\bibnamefont{Chen}},
  \bibinfo{author}{\bibfnamefont{H.}~\bibnamefont{Ren}},
  \bibinfo{author}{\bibfnamefont{F.}~\bibnamefont{Liu}},
  \bibinfo{author}{\bibfnamefont{N.}~\bibnamefont{Li}},
  \bibinfo{author}{\bibfnamefont{Z.}~\bibnamefont{Xing}}, \bibnamefont{et~al.}
  (\bibinfo{year}{2024}), \eprint{arXiv:2311.07353}.

\bibitem[{\citenamefont{Li et~al.}(2024{\natexlab{b}})\citenamefont{Li, Zhang,
  Xiang, Zhang, Zhu, and Wen}}]{li2024signature}
\bibinfo{author}{\bibfnamefont{Q.}~\bibnamefont{Li}},
  \bibinfo{author}{\bibfnamefont{Y.-J.} \bibnamefont{Zhang}},
  \bibinfo{author}{\bibfnamefont{Z.-N.} \bibnamefont{Xiang}},
  \bibinfo{author}{\bibfnamefont{Y.}~\bibnamefont{Zhang}},
  \bibinfo{author}{\bibfnamefont{X.}~\bibnamefont{Zhu}}, \bibnamefont{and}
  \bibinfo{author}{\bibfnamefont{H.-H.} \bibnamefont{Wen}},
  \bibinfo{journal}{Chinese Phys. Lett} \textbf{\bibinfo{volume}{41}},
  \bibinfo{pages}{017401} (\bibinfo{year}{2024}{\natexlab{b}}).

\bibitem[{\citenamefont{Sakakibara
  et~al.}(2024{\natexlab{a}})\citenamefont{Sakakibara, Ochi, Nagata, Ueki,
  Sakurai, Matsumoto, Terashima, Hirose, Ohta, Kato
  et~al.}}]{sakakibara2024theoretical}
\bibinfo{author}{\bibfnamefont{H.}~\bibnamefont{Sakakibara}},
  \bibinfo{author}{\bibfnamefont{M.}~\bibnamefont{Ochi}},
  \bibinfo{author}{\bibfnamefont{H.}~\bibnamefont{Nagata}},
  \bibinfo{author}{\bibfnamefont{Y.}~\bibnamefont{Ueki}},
  \bibinfo{author}{\bibfnamefont{H.}~\bibnamefont{Sakurai}},
  \bibinfo{author}{\bibfnamefont{R.}~\bibnamefont{Matsumoto}},
  \bibinfo{author}{\bibfnamefont{K.}~\bibnamefont{Terashima}},
  \bibinfo{author}{\bibfnamefont{K.}~\bibnamefont{Hirose}},
  \bibinfo{author}{\bibfnamefont{H.}~\bibnamefont{Ohta}},
  \bibinfo{author}{\bibfnamefont{M.}~\bibnamefont{Kato}}, \bibnamefont{et~al.},
  \bibinfo{journal}{Phys. Rev. B} \textbf{\bibinfo{volume}{109}},
  \bibinfo{pages}{144511} (\bibinfo{year}{2024}{\natexlab{a}}).

\bibitem[{\citenamefont{Zhang et~al.}(2024{\natexlab{b}})\citenamefont{Zhang,
  Pei, Du, Hu, Cao, Wang, Wu, Li, Liu, Wen
  et~al.}}]{zhang2024superconductivity}
\bibinfo{author}{\bibfnamefont{M.}~\bibnamefont{Zhang}},
  \bibinfo{author}{\bibfnamefont{C.}~\bibnamefont{Pei}},
  \bibinfo{author}{\bibfnamefont{X.}~\bibnamefont{Du}},
  \bibinfo{author}{\bibfnamefont{W.}~\bibnamefont{Hu}},
  \bibinfo{author}{\bibfnamefont{Y.}~\bibnamefont{Cao}},
  \bibinfo{author}{\bibfnamefont{Q.}~\bibnamefont{Wang}},
  \bibinfo{author}{\bibfnamefont{J.}~\bibnamefont{Wu}},
  \bibinfo{author}{\bibfnamefont{Y.}~\bibnamefont{Li}},
  \bibinfo{author}{\bibfnamefont{H.}~\bibnamefont{Liu}},
  \bibinfo{author}{\bibfnamefont{C.}~\bibnamefont{Wen}}, \bibnamefont{et~al.}
  (\bibinfo{year}{2024}{\natexlab{b}}), \eprint{arXiv:2311.07423}.

\bibitem[{\citenamefont{Li et~al.}(2024{\natexlab{c}})\citenamefont{Li, Chen,
  Huang, Han, Huo, Huang, Ma, Qiu, Chen, Hu et~al.}}]{li2024structural}
\bibinfo{author}{\bibfnamefont{J.}~\bibnamefont{Li}},
  \bibinfo{author}{\bibfnamefont{C.-Q.} \bibnamefont{Chen}},
  \bibinfo{author}{\bibfnamefont{C.}~\bibnamefont{Huang}},
  \bibinfo{author}{\bibfnamefont{Y.}~\bibnamefont{Han}},
  \bibinfo{author}{\bibfnamefont{M.}~\bibnamefont{Huo}},
  \bibinfo{author}{\bibfnamefont{X.}~\bibnamefont{Huang}},
  \bibinfo{author}{\bibfnamefont{P.}~\bibnamefont{Ma}},
  \bibinfo{author}{\bibfnamefont{Z.}~\bibnamefont{Qiu}},
  \bibinfo{author}{\bibfnamefont{J.}~\bibnamefont{Chen}},
  \bibinfo{author}{\bibfnamefont{X.}~\bibnamefont{Hu}}, \bibnamefont{et~al.},
  \bibinfo{journal}{Sci. China: Phys. Mech. Astron.}
  \textbf{\bibinfo{volume}{67}}, \bibinfo{pages}{117403}
  (\bibinfo{year}{2024}{\natexlab{c}}).

\bibitem[{\citenamefont{Huang et~al.}(2024)\citenamefont{Huang, Zhang, Li, Huo,
  Chen, Qiu, Ma, Huang, Sun, and Wang}}]{huang2024signature}
\bibinfo{author}{\bibfnamefont{X.}~\bibnamefont{Huang}},
  \bibinfo{author}{\bibfnamefont{H.}~\bibnamefont{Zhang}},
  \bibinfo{author}{\bibfnamefont{J.}~\bibnamefont{Li}},
  \bibinfo{author}{\bibfnamefont{M.}~\bibnamefont{Huo}},
  \bibinfo{author}{\bibfnamefont{J.}~\bibnamefont{Chen}},
  \bibinfo{author}{\bibfnamefont{Z.}~\bibnamefont{Qiu}},
  \bibinfo{author}{\bibfnamefont{P.}~\bibnamefont{Ma}},
  \bibinfo{author}{\bibfnamefont{C.}~\bibnamefont{Huang}},
  \bibinfo{author}{\bibfnamefont{H.}~\bibnamefont{Sun}}, \bibnamefont{and}
  \bibinfo{author}{\bibfnamefont{M.}~\bibnamefont{Wang}}
  (\bibinfo{year}{2024}), \eprint{arXiv:2410.07861}.

\bibitem[{\citenamefont{Wang et~al.}(2024{\natexlab{b}})\citenamefont{Wang,
  Wen, Wu, Yao, and Xiang}}]{Wang2024CPL}
\bibinfo{author}{\bibfnamefont{M.}~\bibnamefont{Wang}},
  \bibinfo{author}{\bibfnamefont{H.-H.} \bibnamefont{Wen}},
  \bibinfo{author}{\bibfnamefont{T.}~\bibnamefont{Wu}},
  \bibinfo{author}{\bibfnamefont{D.-X.} \bibnamefont{Yao}}, \bibnamefont{and}
  \bibinfo{author}{\bibfnamefont{T.}~\bibnamefont{Xiang}},
  \bibinfo{journal}{Chinese Phys. Lett} \textbf{\bibinfo{volume}{41}},
  \bibinfo{pages}{077402} (\bibinfo{year}{2024}{\natexlab{b}}).

\bibitem[{\citenamefont{Zhang et~al.}(2023)\citenamefont{Zhang, Lin, Moreo,
  Maier, and Dagotto}}]{Zhang2023PRB}
\bibinfo{author}{\bibfnamefont{Y.}~\bibnamefont{Zhang}},
  \bibinfo{author}{\bibfnamefont{L.-F.} \bibnamefont{Lin}},
  \bibinfo{author}{\bibfnamefont{A.}~\bibnamefont{Moreo}},
  \bibinfo{author}{\bibfnamefont{T.~A.} \bibnamefont{Maier}}, \bibnamefont{and}
  \bibinfo{author}{\bibfnamefont{E.}~\bibnamefont{Dagotto}},
  \bibinfo{journal}{Phys. Rev. B} \textbf{\bibinfo{volume}{108}},
  \bibinfo{pages}{165141} (\bibinfo{year}{2023}).

\bibitem[{\citenamefont{Luo et~al.}(2024)\citenamefont{Luo, Lv, Wang, Wú, and
  Yao}}]{Luo2024npj}
\bibinfo{author}{\bibfnamefont{Z.}~\bibnamefont{Luo}},
  \bibinfo{author}{\bibfnamefont{B.}~\bibnamefont{Lv}},
  \bibinfo{author}{\bibfnamefont{M.}~\bibnamefont{Wang}},
  \bibinfo{author}{\bibfnamefont{W.}~\bibnamefont{Wú}}, \bibnamefont{and}
  \bibinfo{author}{\bibfnamefont{D.-X.} \bibnamefont{Yao}},
  \bibinfo{journal}{NPJ QUANTUM MATER} \textbf{\bibinfo{volume}{9}}
  (\bibinfo{year}{2024}), ISSN \bibinfo{issn}{2397-4648}.

\bibitem[{\citenamefont{Yang et~al.}(2023)\citenamefont{Yang, Wang, and
  Wang}}]{Yang2023PRB}
\bibinfo{author}{\bibfnamefont{Q.-G.} \bibnamefont{Yang}},
  \bibinfo{author}{\bibfnamefont{D.}~\bibnamefont{Wang}}, \bibnamefont{and}
  \bibinfo{author}{\bibfnamefont{Q.-H.} \bibnamefont{Wang}},
  \bibinfo{journal}{Phys. Rev. B} \textbf{\bibinfo{volume}{108}},
  \bibinfo{pages}{L140505} (\bibinfo{year}{2023}).

\bibitem[{\citenamefont{Sakakibara
  et~al.}(2024{\natexlab{b}})\citenamefont{Sakakibara, Kitamine, Ochi, and
  Kuroki}}]{Sakakibara2023PRL}
\bibinfo{author}{\bibfnamefont{H.}~\bibnamefont{Sakakibara}},
  \bibinfo{author}{\bibfnamefont{N.}~\bibnamefont{Kitamine}},
  \bibinfo{author}{\bibfnamefont{M.}~\bibnamefont{Ochi}}, \bibnamefont{and}
  \bibinfo{author}{\bibfnamefont{K.}~\bibnamefont{Kuroki}},
  \bibinfo{journal}{Phys. Rev. Lett.} \textbf{\bibinfo{volume}{132}},
  \bibinfo{pages}{106002} (\bibinfo{year}{2024}{\natexlab{b}}).

\bibitem[{\citenamefont{Liu et~al.}(2023{\natexlab{a}})\citenamefont{Liu, Mei,
  Ye, Chen, and Yang}}]{Liu2023PRL}
\bibinfo{author}{\bibfnamefont{Y.-B.} \bibnamefont{Liu}},
  \bibinfo{author}{\bibfnamefont{J.-W.} \bibnamefont{Mei}},
  \bibinfo{author}{\bibfnamefont{F.}~\bibnamefont{Ye}},
  \bibinfo{author}{\bibfnamefont{W.-Q.} \bibnamefont{Chen}}, \bibnamefont{and}
  \bibinfo{author}{\bibfnamefont{F.}~\bibnamefont{Yang}},
  \bibinfo{journal}{Phys. Rev. Lett.} \textbf{\bibinfo{volume}{131}},
  \bibinfo{pages}{236002} (\bibinfo{year}{2023}{\natexlab{a}}).

\bibitem[{\citenamefont{Gu et~al.}(2023)\citenamefont{Gu, Le, Yang, Wu, and
  Hu}}]{gu2023effective}
\bibinfo{author}{\bibfnamefont{Y.}~\bibnamefont{Gu}},
  \bibinfo{author}{\bibfnamefont{C.}~\bibnamefont{Le}},
  \bibinfo{author}{\bibfnamefont{Z.}~\bibnamefont{Yang}},
  \bibinfo{author}{\bibfnamefont{X.}~\bibnamefont{Wu}}, \bibnamefont{and}
  \bibinfo{author}{\bibfnamefont{J.}~\bibnamefont{Hu}} (\bibinfo{year}{2023}),
  \eprint{arXiv:2306.07275}.

\bibitem[{\citenamefont{Qin and Yang}(2023)}]{Qin2023PRB}
\bibinfo{author}{\bibfnamefont{Q.}~\bibnamefont{Qin}} \bibnamefont{and}
  \bibinfo{author}{\bibfnamefont{Y.-f.} \bibnamefont{Yang}},
  \bibinfo{journal}{Phys. Rev. B} \textbf{\bibinfo{volume}{108}},
  \bibinfo{pages}{L140504} (\bibinfo{year}{2023}).

\bibitem[{\citenamefont{Lechermann et~al.}(2023)\citenamefont{Lechermann,
  Gondolf, B\"otzel, and Eremin}}]{Lechermann2023PRB}
\bibinfo{author}{\bibfnamefont{F.}~\bibnamefont{Lechermann}},
  \bibinfo{author}{\bibfnamefont{J.}~\bibnamefont{Gondolf}},
  \bibinfo{author}{\bibfnamefont{S.}~\bibnamefont{B\"otzel}}, \bibnamefont{and}
  \bibinfo{author}{\bibfnamefont{I.~M.} \bibnamefont{Eremin}},
  \bibinfo{journal}{Phys. Rev. B} \textbf{\bibinfo{volume}{108}},
  \bibinfo{pages}{L201121} (\bibinfo{year}{2023}).

\bibitem[{\citenamefont{Heier et~al.}(2024)\citenamefont{Heier, Park, and
  Savrasov}}]{Heier2024PRB}
\bibinfo{author}{\bibfnamefont{G.}~\bibnamefont{Heier}},
  \bibinfo{author}{\bibfnamefont{K.}~\bibnamefont{Park}}, \bibnamefont{and}
  \bibinfo{author}{\bibfnamefont{S.~Y.} \bibnamefont{Savrasov}},
  \bibinfo{journal}{Phys. Rev. B} \textbf{\bibinfo{volume}{109}},
  \bibinfo{pages}{104508} (\bibinfo{year}{2024}).

\bibitem[{\citenamefont{Liu et~al.}(2023{\natexlab{b}})\citenamefont{Liu, Xia,
  Zhou, and Chen}}]{liu2023role}
\bibinfo{author}{\bibfnamefont{H.}~\bibnamefont{Liu}},
  \bibinfo{author}{\bibfnamefont{C.}~\bibnamefont{Xia}},
  \bibinfo{author}{\bibfnamefont{S.}~\bibnamefont{Zhou}}, \bibnamefont{and}
  \bibinfo{author}{\bibfnamefont{H.}~\bibnamefont{Chen}}
  (\bibinfo{year}{2023}{\natexlab{b}}), \eprint{arXiv:2311.07316}.

\bibitem[{\citenamefont{Jiang et~al.}(2024)\citenamefont{Jiang, Wang, and
  Zhang}}]{Jiang2024CPL}
\bibinfo{author}{\bibfnamefont{K.}~\bibnamefont{Jiang}},
  \bibinfo{author}{\bibfnamefont{Z.}~\bibnamefont{Wang}}, \bibnamefont{and}
  \bibinfo{author}{\bibfnamefont{F.-C.} \bibnamefont{Zhang}},
  \bibinfo{journal}{Chinese Physics Letters} \textbf{\bibinfo{volume}{41}},
  \bibinfo{pages}{017402} (\bibinfo{year}{2024}).

\bibitem[{\citenamefont{feng Yang et~al.}(2023)\citenamefont{feng Yang, Zhang,
  and Zhang}}]{yang2023inter}
\bibinfo{author}{\bibfnamefont{Y.}~\bibnamefont{feng Yang}},
  \bibinfo{author}{\bibfnamefont{G.-M.} \bibnamefont{Zhang}}, \bibnamefont{and}
  \bibinfo{author}{\bibfnamefont{F.-C.} \bibnamefont{Zhang}}
  (\bibinfo{year}{2023}), \eprint{arXiv:2308.01176}.

\bibitem[{\citenamefont{Shen et~al.}(2023)\citenamefont{Shen, Qin, and
  Zhang}}]{Shen2023CPL}
\bibinfo{author}{\bibfnamefont{Y.}~\bibnamefont{Shen}},
  \bibinfo{author}{\bibfnamefont{M.}~\bibnamefont{Qin}}, \bibnamefont{and}
  \bibinfo{author}{\bibfnamefont{G.-M.} \bibnamefont{Zhang}},
  \bibinfo{journal}{Chinese Physics Letters} \textbf{\bibinfo{volume}{40}},
  \bibinfo{pages}{127401} (\bibinfo{year}{2023}).

\bibitem[{\citenamefont{Lu et~al.}(2024)\citenamefont{Lu, Pan, Yang, and
  Wu}}]{lu2024interlayer}
\bibinfo{author}{\bibfnamefont{C.}~\bibnamefont{Lu}},
  \bibinfo{author}{\bibfnamefont{Z.}~\bibnamefont{Pan}},
  \bibinfo{author}{\bibfnamefont{F.}~\bibnamefont{Yang}}, \bibnamefont{and}
  \bibinfo{author}{\bibfnamefont{C.}~\bibnamefont{Wu}}, \bibinfo{journal}{Phys.
  Rev. Lett} \textbf{\bibinfo{volume}{132}}, \bibinfo{pages}{146002}
  (\bibinfo{year}{2024}).

\bibitem[{\citenamefont{Qu et~al.}(2024)\citenamefont{Qu, Qu, Chen, Wu, Yang,
  Li, and Su}}]{qu2024PRL}
\bibinfo{author}{\bibfnamefont{X.-Z.} \bibnamefont{Qu}},
  \bibinfo{author}{\bibfnamefont{D.-W.} \bibnamefont{Qu}},
  \bibinfo{author}{\bibfnamefont{J.}~\bibnamefont{Chen}},
  \bibinfo{author}{\bibfnamefont{C.}~\bibnamefont{Wu}},
  \bibinfo{author}{\bibfnamefont{F.}~\bibnamefont{Yang}},
  \bibinfo{author}{\bibfnamefont{W.}~\bibnamefont{Li}}, \bibnamefont{and}
  \bibinfo{author}{\bibfnamefont{G.}~\bibnamefont{Su}}, \bibinfo{journal}{Phys.
  Rev. Lett.} \textbf{\bibinfo{volume}{132}}, \bibinfo{pages}{036502}
  (\bibinfo{year}{2024}).

\bibitem[{\citenamefont{{Fan, Zhen and Zhang, Jian-Feng and Zhan, Bo and Lv,
  Dingshun and Jiang, Xing-Yu and Normand, Bruce and Xiang,
  Tao}}(2024)}]{Fan2024PRB}
\bibinfo{author}{\bibnamefont{{Fan, Zhen and Zhang, Jian-Feng and Zhan, Bo and
  Lv, Dingshun and Jiang, Xing-Yu and Normand, Bruce and Xiang, Tao}}},
  \bibinfo{journal}{Phys. Rev. B} \textbf{\bibinfo{volume}{110}},
  \bibinfo{pages}{024514} (\bibinfo{year}{2024}).

\bibitem[{\citenamefont{Ouyang et~al.}(2024)\citenamefont{Ouyang, Wang, Wang,
  He, Huang, and Lu}}]{Ouyang2024hund}
\bibinfo{author}{\bibfnamefont{Z.}~\bibnamefont{Ouyang}},
  \bibinfo{author}{\bibfnamefont{J.-M.} \bibnamefont{Wang}},
  \bibinfo{author}{\bibfnamefont{J.-X.} \bibnamefont{Wang}},
  \bibinfo{author}{\bibfnamefont{R.-Q.} \bibnamefont{He}},
  \bibinfo{author}{\bibfnamefont{L.}~\bibnamefont{Huang}}, \bibnamefont{and}
  \bibinfo{author}{\bibfnamefont{Z.-Y.} \bibnamefont{Lu}},
  \bibinfo{journal}{Phys. Rev. B} \textbf{\bibinfo{volume}{109}},
  \bibinfo{pages}{115114} (\bibinfo{year}{2024}).

\bibitem[{\citenamefont{Luo et~al.}(2023)\citenamefont{Luo, Hu, Wang, W{\'u},
  and Yao}}]{luo2023bilayer}
\bibinfo{author}{\bibfnamefont{Z.}~\bibnamefont{Luo}},
  \bibinfo{author}{\bibfnamefont{X.}~\bibnamefont{Hu}},
  \bibinfo{author}{\bibfnamefont{M.}~\bibnamefont{Wang}},
  \bibinfo{author}{\bibfnamefont{W.}~\bibnamefont{W{\'u}}}, \bibnamefont{and}
  \bibinfo{author}{\bibfnamefont{D.-X.} \bibnamefont{Yao}},
  \bibinfo{journal}{Phys. Rev. Lett} \textbf{\bibinfo{volume}{131}},
  \bibinfo{pages}{126001} (\bibinfo{year}{2023}).

\bibitem[{\citenamefont{Taniguchi et~al.}(1995)\citenamefont{Taniguchi,
  Nishikawa, Yasui, Kobayashi, Takeda, Shamoto, and
  Sato}}]{taniguchi1995transport}
\bibinfo{author}{\bibfnamefont{S.}~\bibnamefont{Taniguchi}},
  \bibinfo{author}{\bibfnamefont{T.}~\bibnamefont{Nishikawa}},
  \bibinfo{author}{\bibfnamefont{Y.}~\bibnamefont{Yasui}},
  \bibinfo{author}{\bibfnamefont{Y.}~\bibnamefont{Kobayashi}},
  \bibinfo{author}{\bibfnamefont{J.}~\bibnamefont{Takeda}},
  \bibinfo{author}{\bibfnamefont{S.-i.} \bibnamefont{Shamoto}},
  \bibnamefont{and} \bibinfo{author}{\bibfnamefont{M.}~\bibnamefont{Sato}},
  \bibinfo{journal}{J. Phys. Soc. Jpn.} \textbf{\bibinfo{volume}{64}},
  \bibinfo{pages}{1644} (\bibinfo{year}{1995}).

\bibitem[{\citenamefont{Dong et~al.}(2024)\citenamefont{Dong, Huo, Li, Li, Li,
  Sun, Gu, Lu, Wang, Wang et~al.}}]{Dong2024Nature}
\bibinfo{author}{\bibfnamefont{Z.}~\bibnamefont{Dong}},
  \bibinfo{author}{\bibfnamefont{M.}~\bibnamefont{Huo}},
  \bibinfo{author}{\bibfnamefont{J.}~\bibnamefont{Li}},
  \bibinfo{author}{\bibfnamefont{J.}~\bibnamefont{Li}},
  \bibinfo{author}{\bibfnamefont{P.}~\bibnamefont{Li}},
  \bibinfo{author}{\bibfnamefont{H.}~\bibnamefont{Sun}},
  \bibinfo{author}{\bibfnamefont{L.}~\bibnamefont{Gu}},
  \bibinfo{author}{\bibfnamefont{Y.}~\bibnamefont{Lu}},
  \bibinfo{author}{\bibfnamefont{M.}~\bibnamefont{Wang}},
  \bibinfo{author}{\bibfnamefont{Y.}~\bibnamefont{Wang}}, \bibnamefont{et~al.},
  \bibinfo{journal}{Nature} \textbf{\bibinfo{volume}{630}},
  \bibinfo{pages}{847–852} (\bibinfo{year}{2024}), ISSN
  \bibinfo{issn}{1476-4687}.

\bibitem[{\citenamefont{Zhang et~al.}(2024{\natexlab{c}})\citenamefont{Zhang,
  Lin, Moreo, Maier, and Dagotto}}]{Zhang2024PRB}
\bibinfo{author}{\bibfnamefont{Y.}~\bibnamefont{Zhang}},
  \bibinfo{author}{\bibfnamefont{L.-F.} \bibnamefont{Lin}},
  \bibinfo{author}{\bibfnamefont{A.}~\bibnamefont{Moreo}},
  \bibinfo{author}{\bibfnamefont{T.~A.} \bibnamefont{Maier}}, \bibnamefont{and}
  \bibinfo{author}{\bibfnamefont{E.}~\bibnamefont{Dagotto}},
  \bibinfo{journal}{Phys. Rev. B} \textbf{\bibinfo{volume}{109}},
  \bibinfo{pages}{045151} (\bibinfo{year}{2024}{\natexlab{c}}).

\bibitem[{\citenamefont{Cui et~al.}(2023)\citenamefont{Cui, Choi, Lin, Liu,
  Wang, Wang, Chen, Hong, Rong, Wang et~al.}}]{cui2023strain}
\bibinfo{author}{\bibfnamefont{T.}~\bibnamefont{Cui}},
  \bibinfo{author}{\bibfnamefont{S.}~\bibnamefont{Choi}},
  \bibinfo{author}{\bibfnamefont{T.}~\bibnamefont{Lin}},
  \bibinfo{author}{\bibfnamefont{C.}~\bibnamefont{Liu}},
  \bibinfo{author}{\bibfnamefont{G.}~\bibnamefont{Wang}},
  \bibinfo{author}{\bibfnamefont{N.}~\bibnamefont{Wang}},
  \bibinfo{author}{\bibfnamefont{S.}~\bibnamefont{Chen}},
  \bibinfo{author}{\bibfnamefont{H.}~\bibnamefont{Hong}},
  \bibinfo{author}{\bibfnamefont{D.}~\bibnamefont{Rong}},
  \bibinfo{author}{\bibfnamefont{Q.}~\bibnamefont{Wang}}, \bibnamefont{et~al.}
  (\bibinfo{year}{2023}), \eprint{arXiv:2311.13228}.

\bibitem[{\citenamefont{Zhang and Greenblatt}(1994)}]{zhang1994structure}
\bibinfo{author}{\bibfnamefont{Z.}~\bibnamefont{Zhang}} \bibnamefont{and}
  \bibinfo{author}{\bibfnamefont{M.}~\bibnamefont{Greenblatt}},
  \bibinfo{journal}{J. Solid. State Chem (United States)}
  \textbf{\bibinfo{volume}{111}} (\bibinfo{year}{1994}).

\bibitem[{\citenamefont{Xu et~al.}(2023)\citenamefont{Xu, Huyan, Wang, Bud'ko,
  Chen, Ke, Mitchell, Canfield, Li, and Xie}}]{xu2023pressure}
\bibinfo{author}{\bibfnamefont{M.}~\bibnamefont{Xu}},
  \bibinfo{author}{\bibfnamefont{S.}~\bibnamefont{Huyan}},
  \bibinfo{author}{\bibfnamefont{H.}~\bibnamefont{Wang}},
  \bibinfo{author}{\bibfnamefont{S.~L.} \bibnamefont{Bud'ko}},
  \bibinfo{author}{\bibfnamefont{X.}~\bibnamefont{Chen}},
  \bibinfo{author}{\bibfnamefont{X.}~\bibnamefont{Ke}},
  \bibinfo{author}{\bibfnamefont{J.~F.} \bibnamefont{Mitchell}},
  \bibinfo{author}{\bibfnamefont{P.~C.} \bibnamefont{Canfield}},
  \bibinfo{author}{\bibfnamefont{J.}~\bibnamefont{Li}}, \bibnamefont{and}
  \bibinfo{author}{\bibfnamefont{W.}~\bibnamefont{Xie}} (\bibinfo{year}{2023}),
  \eprint{arXiv:2312.14251}.

\bibitem[{\citenamefont{Jiao et~al.}(2024)\citenamefont{Jiao, Niu, Xu, Zhen,
  Wang, and Zhang}}]{Jiao20241354504}
\bibinfo{author}{\bibfnamefont{K.}~\bibnamefont{Jiao}},
  \bibinfo{author}{\bibfnamefont{R.}~\bibnamefont{Niu}},
  \bibinfo{author}{\bibfnamefont{H.}~\bibnamefont{Xu}},
  \bibinfo{author}{\bibfnamefont{W.}~\bibnamefont{Zhen}},
  \bibinfo{author}{\bibfnamefont{J.}~\bibnamefont{Wang}}, \bibnamefont{and}
  \bibinfo{author}{\bibfnamefont{C.}~\bibnamefont{Zhang}},
  \bibinfo{journal}{Physica C} \textbf{\bibinfo{volume}{621}},
  \bibinfo{pages}{1354504} (\bibinfo{year}{2024}).

\bibitem[{\citenamefont{Poltavets et~al.}(2006)\citenamefont{Poltavets,
  Lokshin, Egami, and Greenblatt}}]{poltavets2006oxygen}
\bibinfo{author}{\bibfnamefont{V.~V.} \bibnamefont{Poltavets}},
  \bibinfo{author}{\bibfnamefont{K.~A.} \bibnamefont{Lokshin}},
  \bibinfo{author}{\bibfnamefont{T.}~\bibnamefont{Egami}}, \bibnamefont{and}
  \bibinfo{author}{\bibfnamefont{M.}~\bibnamefont{Greenblatt}},
  \bibinfo{journal}{Mater. Res. Bull.} \textbf{\bibinfo{volume}{41}},
  \bibinfo{pages}{955} (\bibinfo{year}{2006}).

\bibitem[{\citenamefont{Li et~al.}(2020)\citenamefont{Li, Guo, Zhang, Song,
  Gao, Gu, and Nie}}]{li2020epitaxial}
\bibinfo{author}{\bibfnamefont{Z.}~\bibnamefont{Li}},
  \bibinfo{author}{\bibfnamefont{W.}~\bibnamefont{Guo}},
  \bibinfo{author}{\bibfnamefont{T.}~\bibnamefont{Zhang}},
  \bibinfo{author}{\bibfnamefont{J.}~\bibnamefont{Song}},
  \bibinfo{author}{\bibfnamefont{T.}~\bibnamefont{Gao}},
  \bibinfo{author}{\bibfnamefont{Z.}~\bibnamefont{Gu}}, \bibnamefont{and}
  \bibinfo{author}{\bibfnamefont{Y.}~\bibnamefont{Nie}}, \bibinfo{journal}{APL
  Mater} \textbf{\bibinfo{volume}{8}} (\bibinfo{year}{2020}).

\bibitem[{\citenamefont{Liu et~al.}(2024)\citenamefont{Liu, Ou, Chu, Yang, Li,
  Zhang, and Wen}}]{liu2024growth}
\bibinfo{author}{\bibfnamefont{Y.}~\bibnamefont{Liu}},
  \bibinfo{author}{\bibfnamefont{M.}~\bibnamefont{Ou}},
  \bibinfo{author}{\bibfnamefont{H.}~\bibnamefont{Chu}},
  \bibinfo{author}{\bibfnamefont{H.}~\bibnamefont{Yang}},
  \bibinfo{author}{\bibfnamefont{Q.}~\bibnamefont{Li}},
  \bibinfo{author}{\bibfnamefont{Y.}~\bibnamefont{Zhang}}, \bibnamefont{and}
  \bibinfo{author}{\bibfnamefont{H.-H.} \bibnamefont{Wen}}
  (\bibinfo{year}{2024}), \eprint{arXiv:2406.08789}.

\bibitem[{\citenamefont{Kiselev et~al.}(2019)\citenamefont{Kiselev, Gaczynski,
  Eckold, Feldhoff, Becker, and Cherepanov}}]{kiselev2019investigations}
\bibinfo{author}{\bibfnamefont{E.~A.} \bibnamefont{Kiselev}},
  \bibinfo{author}{\bibfnamefont{P.}~\bibnamefont{Gaczynski}},
  \bibinfo{author}{\bibfnamefont{G.}~\bibnamefont{Eckold}},
  \bibinfo{author}{\bibfnamefont{A.}~\bibnamefont{Feldhoff}},
  \bibinfo{author}{\bibfnamefont{K.-D.} \bibnamefont{Becker}},
  \bibnamefont{and} \bibinfo{author}{\bibfnamefont{V.~A.}
  \bibnamefont{Cherepanov}}, \bibinfo{journal}{Chimica Techno Acta. 2019. Vol.
  6.№ 2} \textbf{\bibinfo{volume}{6}}, \bibinfo{pages}{51}
  (\bibinfo{year}{2019}).

\bibitem[{\citenamefont{Zhou et~al.}(2024{\natexlab{b}})\citenamefont{Zhou,
  Guo, Cai, Sun, Wang, Zhao, Han, Chen, Chen, Wu et~al.}}]{zhou2024evidence}
\bibinfo{author}{\bibfnamefont{Y.}~\bibnamefont{Zhou}},
  \bibinfo{author}{\bibfnamefont{J.}~\bibnamefont{Guo}},
  \bibinfo{author}{\bibfnamefont{S.}~\bibnamefont{Cai}},
  \bibinfo{author}{\bibfnamefont{H.}~\bibnamefont{Sun}},
  \bibinfo{author}{\bibfnamefont{P.}~\bibnamefont{Wang}},
  \bibinfo{author}{\bibfnamefont{J.}~\bibnamefont{Zhao}},
  \bibinfo{author}{\bibfnamefont{J.}~\bibnamefont{Han}},
  \bibinfo{author}{\bibfnamefont{X.}~\bibnamefont{Chen}},
  \bibinfo{author}{\bibfnamefont{Y.}~\bibnamefont{Chen}},
  \bibinfo{author}{\bibfnamefont{Q.}~\bibnamefont{Wu}}, \bibnamefont{et~al.}
  (\bibinfo{year}{2024}{\natexlab{b}}), \eprint{arXiv:2311.12361}.

\bibitem[{\citenamefont{Wú et~al.}(2024)\citenamefont{Wú, Luo, Yao, and
  Wang}}]{Wu2024superexchange}
\bibinfo{author}{\bibfnamefont{W.}~\bibnamefont{Wú}},
  \bibinfo{author}{\bibfnamefont{Z.}~\bibnamefont{Luo}},
  \bibinfo{author}{\bibfnamefont{D.-X.} \bibnamefont{Yao}}, \bibnamefont{and}
  \bibinfo{author}{\bibfnamefont{M.}~\bibnamefont{Wang}}, \bibinfo{journal}{SCI
  CHINA PHYS MECH} \textbf{\bibinfo{volume}{67}} (\bibinfo{year}{2024}), ISSN
  \bibinfo{issn}{1869-1927}.

\bibitem[{\citenamefont{Ziman}(2001)}]{ziman2001electrons}
\bibinfo{author}{\bibfnamefont{J.~M.} \bibnamefont{Ziman}},
  \emph{\bibinfo{title}{{Electrons and phonons: the theory of transport
  phenomena in solids}}} (\bibinfo{publisher}{Oxford university press},
  \bibinfo{year}{2001}).

\end{thebibliography}
\end{document}